\documentclass[10pt,aps,prb,twocolumn,superscriptaddress,floatfix,showpacs,longbibliography]{revtex4-1}
\usepackage[utf8]{inputenc}
\usepackage{graphicx}
\usepackage{tabularx}
\usepackage[usenames,dvipsnames,table]{xcolor}
\usepackage[version=3]{mhchem}
\usepackage{braket}
\usepackage{upgreek}
\usepackage{hyperref}
\usepackage{amsmath, amssymb}
\usepackage[normalem]{ulem}
\usepackage{soul}

\definecolor{mark}{rgb}{0.85, 0.9, 1}
\definecolor{rred}{HTML}{CB4154}

\hypersetup{hidelinks}
\sethlcolor{mark}
\newcolumntype{Y}{>{\centering\arraybackslash}X}

\begin{document}

\title{Quantifying spatio-temporal patterns in classical and quantum systems out of equilibrium}
\author{E. A. Maletskii$^{1}$, I. A. Iakovlev$^{1}$, V.~V.~Mazurenko}

\affiliation{ Theoretical Physics and Applied Mathematics Department, Ural Federal University, Mira Str. 19, 620002 Ekaterinburg, Russia}

\date{\today}

\begin{abstract}
A rich variety of non-equilibrium dynamical phenomena and processes unambiguously calls for the development of general numerical techniques to probe and estimate a complex interplay between spatial and temporal degrees of freedom in many-body systems of completely different nature. In this work we provide a solution to this problem by adopting a structural complexity measure to quantify spatio-temporal patterns in the time-dependent digital representation of a system. On the basis of very limited amount of data our approach allows to distinguish different dynamical regimes and define critical parameters in both classical and quantum systems. By the example of the discrete time crystal realized in non-equilibrium quantum systems we provide a complete low-level characterization of this nontrivial dynamical phase with only processing bitstrings, which can be considered as a valuable alternative to previous studies based on the calculations of qubit correlation functions.
\end{abstract}

\maketitle

\section*{Introduction}
The world around us is constantly changing, which makes dynamics of the artificial and natural systems at different spatial scales is of great interest to the scientific community. Traditionally, to describe a dynamical system one explores a time-dependent behaviour of a relevant physical quantity. In certain situations the choice of such a quantity is obvious, for instance description of a planetary motion is fulfilled by analyzing the trajectories, time-varying coordinates of the planets. On the other hand, in the case of the systems out of equilibrium the search for a suitable quantity could be a non-trivial task. Here, a bright primer known from the school textbooks is a microparticle in a liquid environment characterized by Brownian motion \cite{Brown}. It was found that observations over the trajectory of an individual particle does not allow to characterize the random motion, the only time-dependent mean-square displacement \cite{Einstein, Langevin} or velocity autocorrelation function averaged over many particles trajectories matter. Importantly, this fundamental finding has determined a general strategy for the analysis of non-equilibrium motion in physical systems of one or a few bodies within the standard statistical approaches or advanced machine learning techniques\cite{Granik, ACSnano}. 

\begin{figure}[!hb]
		\centering
		\includegraphics[width=0.45\textwidth]{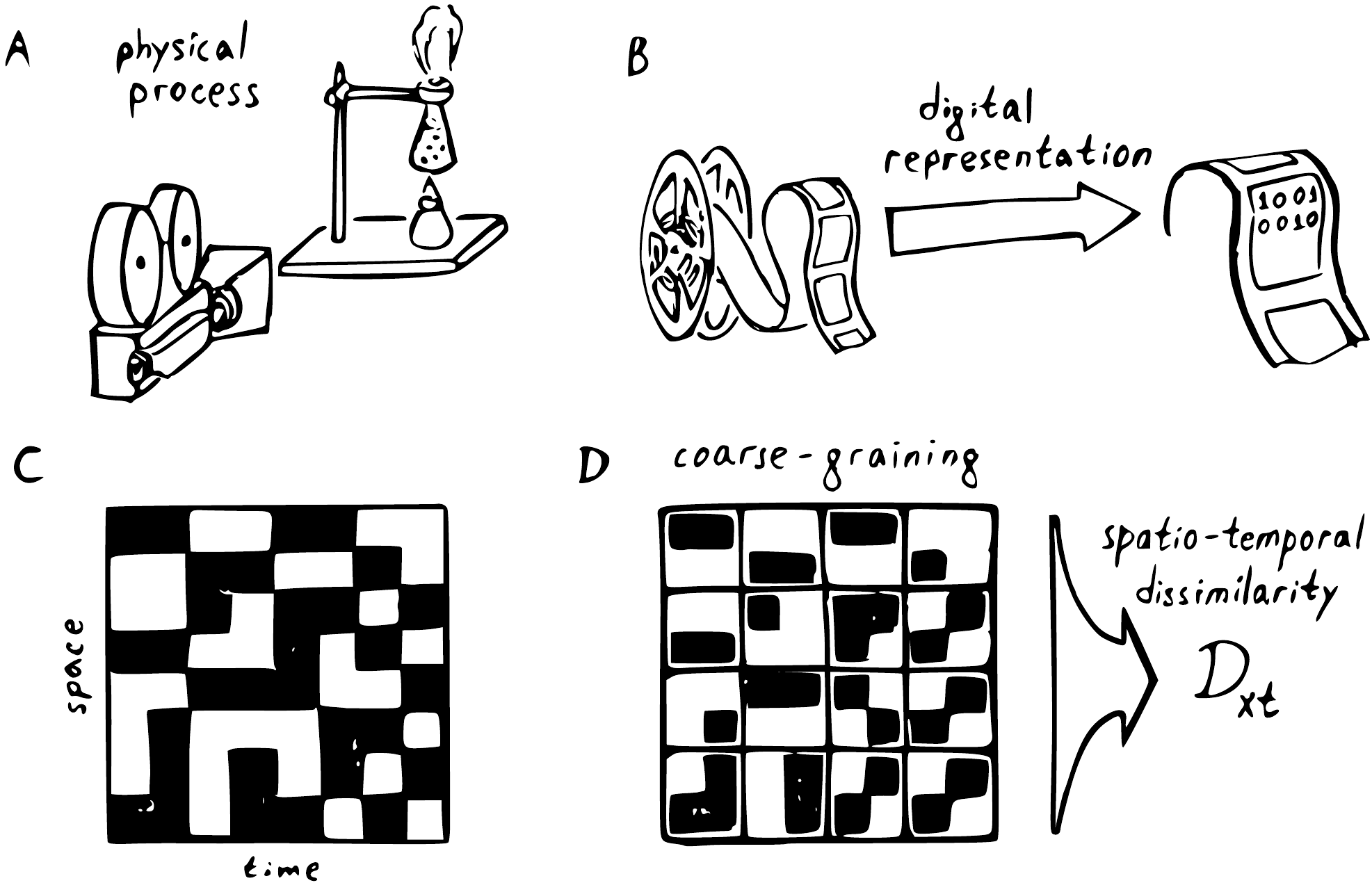}
		\caption{\label{fig1} Pictorial representation of the protocol realizing the concept one process - one number. A physical process of our interest is recorded (A) and then is converted into a set of frames with one-dimensional digital representations (B) of the system at each time moment. (C) These frames are merged in a space-time grid. (D) By using the renormalization procedure with spatio-temporal filters discussed in the text we attribute the initial physical process to a single number (spatio-temporal dissimilarity) characterzing the dissimilarity of the patterns in the digital system's representation on the space-time grid.}
	\end{figure}

A wide diversity of non-equilibrium processes in classical systems stimulates permanent search for distinct measures \cite{compression1, compression2, Falk} that allow  theoretical characterization of the observed dynamics. Of particular difficulty is the description and prediction of a collective behavior of classical systems consisting of many interacting and simultaneously moving elements\cite{protein, colloid} or participants\cite{flocks, sheeps, pedestrian1,pedestrian2}. The standard choice here is the calculations of correlation functions that are non-local in both time and space for some physical quantities characterized the system in question. Depending on the situation these quantities can be taken directly from the experiment as they are (for instance, time-dependent positions of the particles obtained with a tracking system \cite{colloid,Scalliet}), or one could define some useful features \cite{Gnesotto} on the basis of available experimental data.  The problem is that even when defining the spatio-temporal correlators there is no a universal method for extracting a meaningful information from them, for instance, a signature of a hidden dynamical order. One of the possible solutions to simplify the consideration is to decouple the intertwined spatial and temporal correlations and explore them separately \cite{flocks, sheeps}. It means that potentially important information on dynamics of the system in question may be lost. In certain situations the choice of measures taken from information theory and related to entropy or complexity gives very promising results \cite{fallpaper1, fallpaper2, Mutual_info, Palmer} in solving the characterization task.  

Non-equilibrium physics is not limited to the classical case. For instance, when describing physical properties of quantum lattice  systems one faces the challenging problem of calculating the quantum-mechanical time evolution of many-body systems out of equilibrium, which can be done with the nonequilibrium extension of the dynamical mean field theory \cite{Werner1, Werner2}.
Besides, significant theoretical and experimental efforts are concentrated around quantum matter and phenomena out of equilibrium for which none classical analogues exist. They include exploration of thermalization dynamics \cite{thermalization1}, hydrodynamic transport \cite{transport, transport2, transport3},  quantum information scrambling \cite{scrabmling_Google}, non-equilibrium phases that are formed in periodically-driven quantum systems \cite{CNN_MBL, DTC_IBM, DTC_Google}, topological order emerge in out-of-equilibrium states of programmable Rydberg atom arrays \cite{Lukin1}, quantum chaos \cite{quantum_chaos} and other non-trivial findings that play a crucial role in further developing quantum technologies, information theory and condensed matter physics.  In this regard, quantum simulators  and quantum computers facilitate the modeling of the non-equilibrium quantum dynamics \cite{Google_Majorana} and represent a promising alternative to classical supercomputers whose using is fundamentally limited due to the exponential growth of the Hilbert space. Similar to the classical case, theoretical analysis of non-equilibrium quantum systems can be simplified by calculating temporal autocorrelators that are local in space and the spatial correlation functions that are local in time. However, such a separate consideration of the spatial and temporal correlations inevitably leads to loss of information on interplay between space and time degrees of freedom.

In this paper we propose an alternative approach for exploring the non-equilibrium dynamics of classical and quantum systems by considering space and time degrees of freedom simultaneously and, thus, without uncoupling spatio-temporal correlations that could contain important information on the state of the dynamical system. To this aim, we follow the idea one process - one number which is schematically visualized in Fig.\ref{fig1}. It includes recording a process at given external conditions (Fig.\ref{fig1} A), converting the recorded movie into digital representation as a spatio-temporal image (Figs.\ref{fig1} B and C) and estimating patterns of the image by means of the structural complexity (dissimilarity) measure recently introduced in Refs.\cite{PNASComplexity, Dissimilarity} (Fig.\ref{fig1} D). As we will show by classical and quantum examples such a consideration of the dynamical process eases the search for critical values of the external parameters at which there are global changes in the dynamical properties of the physical system in question. Generality and flexibility in exploring completely different non-equilibrium quantum or classical systems within the same framework are important features of the proposed scheme. For instance, we perform a characterization of discrete time crystal - a complex out-of-equilibrium phase of the quantum matter by means of the spatio-temporal dissimilarity introduced in this investigation and some measures taken from the information theory. It shows that the theoretical description of non-equilibrium dynamical processes without calculating correlation functions is possible.  

\section*{Methods}

Our approach to the analysis of the quantum or classical dynamics out of equilibrium is formulated on the basis of the recently proposed renormalization group (RG) algorithm \cite{PNASComplexity} for estimating the structural complexity of an object. The essence of this method is to quantify dissimilarities between the patterns that occur at different spatial scales.  Such an approach was found extremely useful in context of defining equilibrium  phase transitions of various nature~\cite{PNASComplexity, Khatami} as well as verification of quantum wave functions \cite{Dissimilarity}. Since in quantum case this approach is hardly related to the complexity of a quantum state itself and operate with the array of single measurements (bitstrings) the term dissimilarity was introduced to name the procedure\cite{Dissimilarity}. 
    
In this paper, to characterize a dynamical process we propose the spatio-temporal dissimilarity measure. The idea is based on dealing with two-dimensional patterns that occur when one puts the digital representations of the system states at different steps of its time evolution together. Having a one-dimensional discrete system of size $L$ and its evolution over $T$ time steps we can construct a matrix $B^0_{L\times T}$. At every step $k>0$ of a coarse-graining procedure we create a matrix $B^k_{L\times T}$ of the same size, whose elements are calculated as follows
    \begin{equation}\label{Coarse-graining procedure}
        b^k_{ij} = \frac{1}{\Lambda^k_x\Lambda^k_t} \sum_{l=1}^{\Lambda^k_x} \sum_{m=1}^{\Lambda^k_t} b^{k-1}_{\Lambda^k_x [(i-1)/\Lambda^k_x]+l, \Lambda^k_t[(j-1)/\Lambda^k_t]+m},
    \end{equation}
where square brackets represent taking integer part and $l$($m$) indexes denote elements within a matrix block of $\Lambda^k_x\times\Lambda^k_t$. 
Thus, the whole matrix is divided into blocks of $\Lambda^k_x\times\Lambda^k_t$ size (in this particular case, $k$ is a degree, not an index) and the elements within each block are substituted with its average value. 

In this work $\Lambda_{\rm x} = \Lambda_{\rm t} = 2$ space-time filters were used, since, as we will show below, they provide a complete description of all the considered cases, both classical and quantum. However, for arbitrary non-equilibrium system one needs to test different spatio-temporal filters to be sure on extracting maximal useful information with dissimilarity measure. For convenience, elements of the initial matrix are normalized to the range $[-1;1]$.

After that we compute the partial spatio-temporal dissimilarities for the patterns separated by one step of RG. In case of simple averaging scheme used in our calculations, this can be done using the following equation
    
   \begin{equation}\label{D_k}
		\mathcal{D}_{\rm xt}^k = \frac{1}{2LT}\left|\left(\sum_{i,j=1}^{L,T} \left[(b^{k+1}_{ij})^2 - (b^{k}_{ij})^2 \right]\right) \right|.
   \end{equation}
	
	Finally, given features emerging at every new scale, we obtain
	\begin{equation}\label{Complexity}
		\mathcal{D}_{\rm xt} = \sum_{k=1}^{N}\mathcal{D}_{\rm xt}^k ,
	\end{equation} where $N$ is the total number of RG steps.  Thus, within this protocol each dynamical process can be associated to a single number characterizing the dissimilarity of the system's patterns in the spatio-temporal grid.
 The described procedure is then technically identical to the calculation of the structural complexity of the two-dimensional grayscale image \cite{PNASComplexity}.  

An independent choice of the initial filter sizes ($\Lambda_{\rm x}$ and $\Lambda_{\rm t}$) allows a selective quantifying either spatial or temporal patterns. For instance, if one chooses $\Lambda_{\rm x} = 1$ and $\Lambda_{\rm t} > 1$, then $\mathcal{D}_{\rm xt}$ boils down to $\mathcal{D}_{\rm t}$ that estimates temporal patterns individually for each spatial degree of freedom. On the other hand, for $\Lambda_{\rm x} > 1$ and $\Lambda_{\rm t} = 1$, one gets $\mathcal{D}_{\rm x}$ that probes spatial patterns in the non-equilibrium system. The spatial (temporal) dissimilarity of the whole dynamical process is then taken as the average value of the elements of the corresponding $1\times T$ ($L\times 1$) vector. 

Thus, the complete set of dissimilarity functions, $\mathcal{D}_{\rm xt}$, $\mathcal{D}_{\rm t}$ and $\mathcal{D}_{\rm x}$ including their partial contributions can be used for a low-level characterization of non-equilibrium dynamics in a quantum 
 or classical system without searching for quantities that contain basic information on the system and calculating correlation functions for them.

\section*{Results}
\subsection*{1D lattice gas model}

\begin{figure}[t]
\centering
\includegraphics[width=0.75\columnwidth]{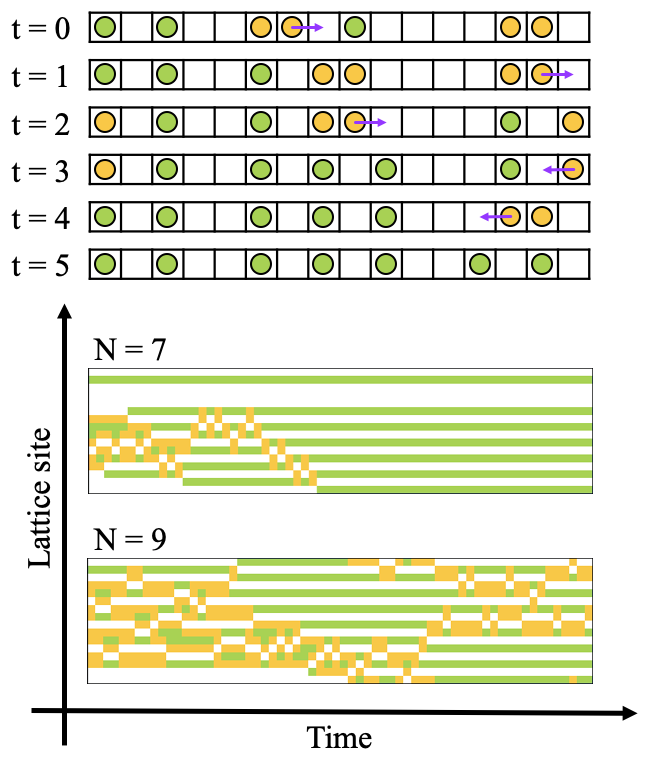}
\caption{\label{fig:clg_model} (Top) Schematic representation of the CLG model on a chain with periodic boundaries. At each time step one randomly chosen active particle (depicted in orange) moves to the unoccupied position. The dynamic stops when there are no active particles (t = 5). (Bottom) Examples of the time evolution of the L = 16 chain with N = 7 (adsorbing state) and N = 9 (active state) occupied sites.}
\end{figure}

\begin{figure}[t]
\centering
\includegraphics[width=0.8\columnwidth]{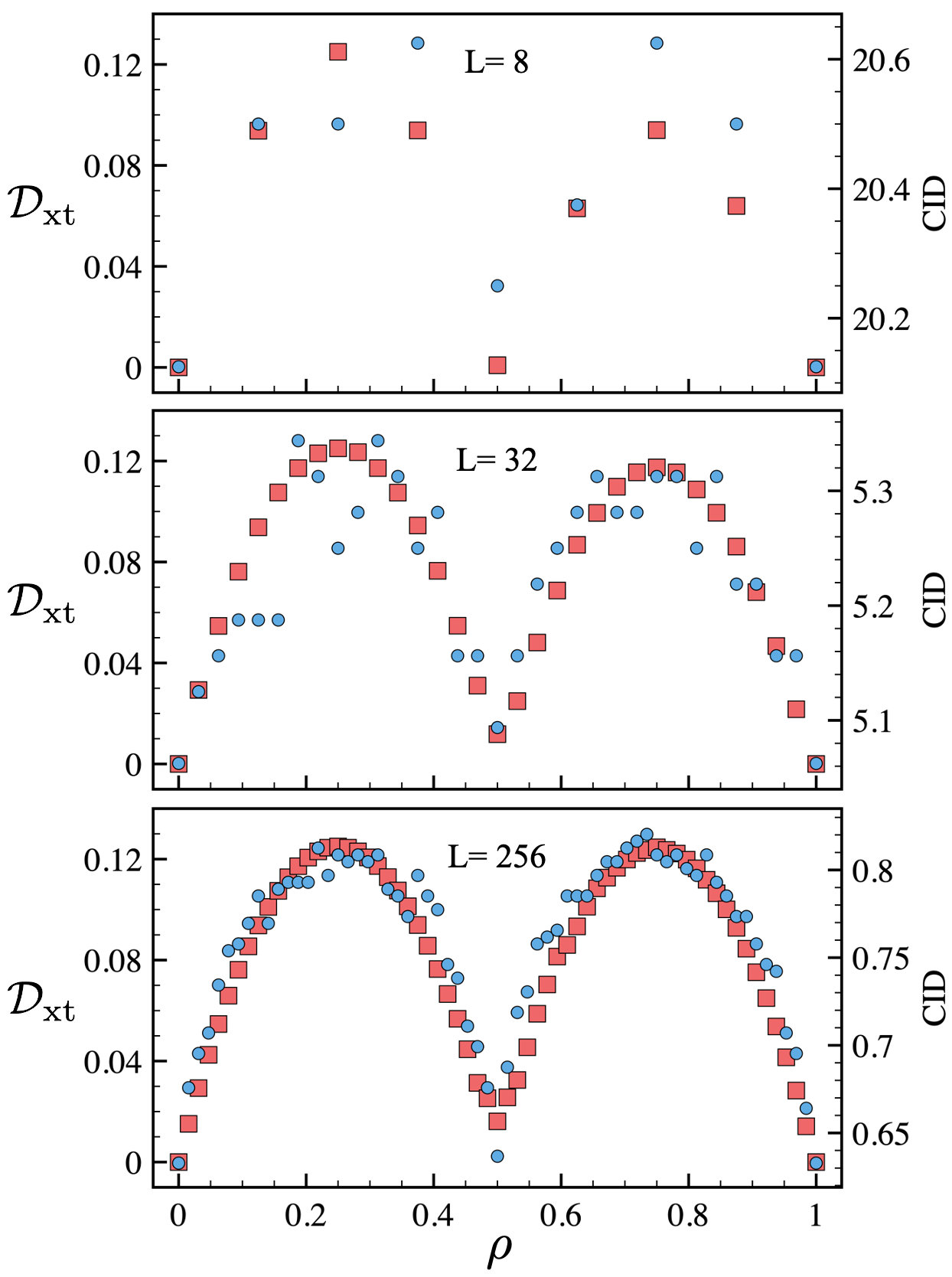}
\caption{\label{fig:1d_gas} Comparison of the $\mathcal{D}_{\rm xt}$ (red squares) and CID (blue circles) behaviour in case of CLG problem on 1D chains of different size.}
\end{figure}

To assess the viability of the spatio-temporal dissimilarity approach we first consider a simple conserved lattice gas (CLG) model in 1D~\cite{clg1, clg2} which is one of the sandpile models to study self-organised criticality. The initial state of the system in question is a random distribution of $N$ occupied sites on the $L\ge N$ chain. Particles, for which one of its neighbouring site is occupied are considered as active. Dynamics in the CLG model can be described in the following way: at each time step one should randomly choose an active particle and move it to the unoccupied neighbouring position. For $\rho = N/L \le 0.5$ the system ends up in a steady adsorbing state without active sites meanwhile for higher densities the dynamical fluctuations will take place at any time. Thus, we have a nonequilibrium phase transition with the order parameter $f_a$ denoting a fraction of active states on the final lattice configuration. Schematic representation of the CLG model as well as the examples of real time evolution of the system belonging to different phases are given in Fig.~\ref{fig:clg_model}. 

The authors of Ref.~\cite{clg_cid} have shown, that the phase transition between the adsorbing and active states in this model can be easily defined in the thermodynamic limit with using a computable information density (CID). To calculate this quantity one needs to store the final state of the chain in the data file $F$ and then compress it without the losses of information to get $F^\prime$. CID is simply equal to the ratio between the sizes of $F^\prime$ and $F$ in bytes. The idea is based on the intuitive assumption that the more structured the file, the less information is needed to reproduce its content, which is similar to the concept of Kolmogorov complexity~\cite{Kolmogorov}. It was demonstrated that the method facilitates accurate description of the thermodynamic phase transitions in equilibrium magnetic systems\cite{compression2}, reveals hidden dynamical ordering in classical systems out of equilibrium \cite{clg_cid} and helps to solve other interesting problems \cite{compression4, compression5}. Importantly, the compression algorithm works effectively if two criteria are met: one has the large amount of data, and the data itself fits a dictionary with discrete set of items.

\begin{figure}[t]
\centering
\includegraphics[width=1.0\columnwidth]{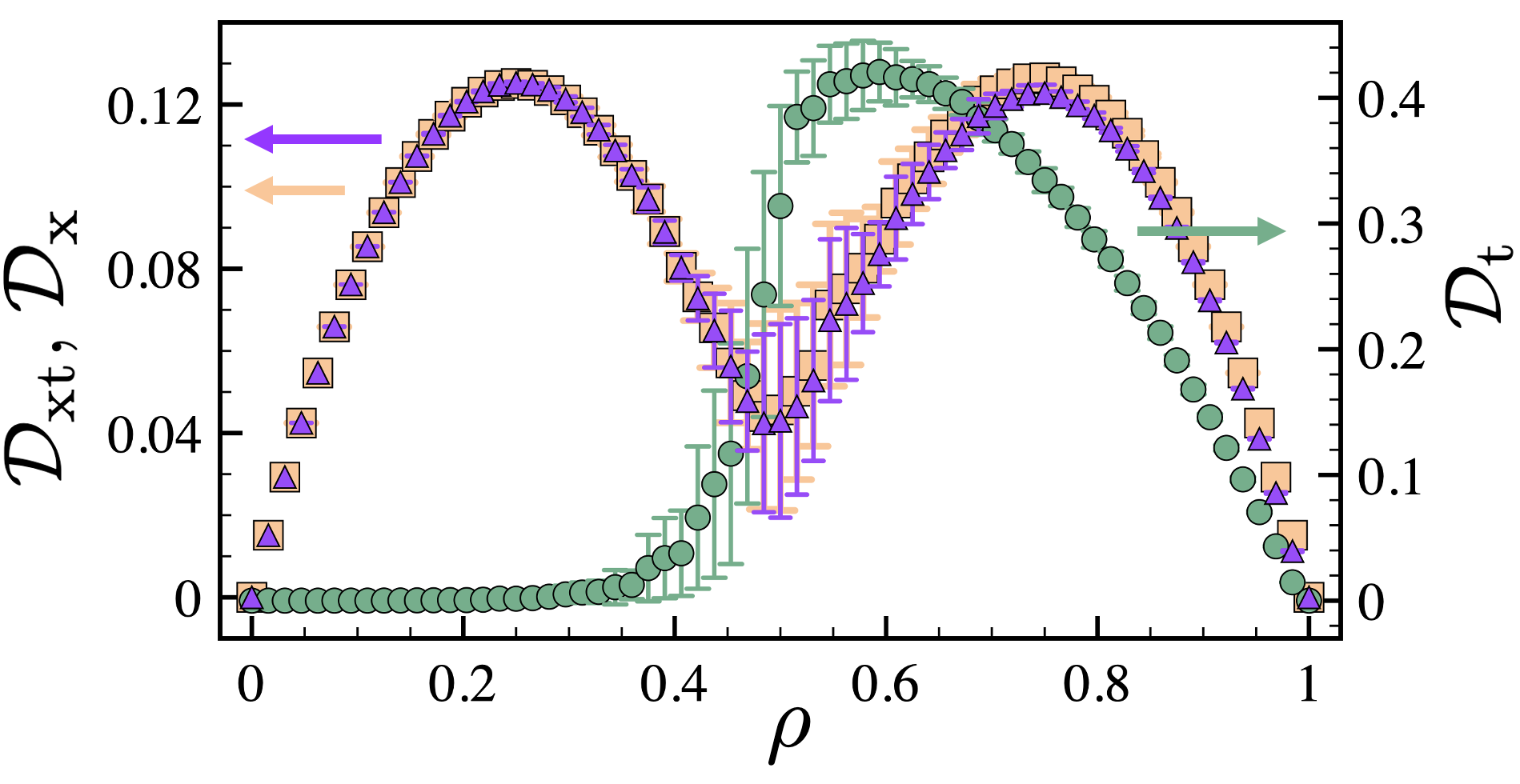}
\caption{\label{fig:d_avg} Comparison of the $\mathcal{D}_{\rm xt}$ (purple triangles), $\mathcal{D}_{\rm x}$ (orange squares) and $\mathcal{D}_{\rm t}$ (green circles) behaviour in case of CLG problem on 1D chain with $L=64$, $T=1024$. The results are averaged over 100 independent starting configurations at each density.}
\end{figure}

Here we compare $\mathcal{D}_{\rm xt}$ with CID in case of small chains with up to $L = 256$ sites. As can be seen from Fig.~\ref{fig:1d_gas}, the maximum value of dissimilarity is not sensitive to the size of the considered system. In turn, CID is not only exhibit strong fluctuations, but also lies in a significantly different range depending on $L$. Here we use Lempel-Ziv-Markov chain-Algorithm (LZMA) as it is one of the extensions of Lempel-Ziv 77 (LZ77) compression algorithm used in Ref.~\cite{clg_cid} and shows the best results among other standard techniques. When decreasing the system size the amplitude of $\mathcal{D}_{\rm xt}$ drops a bit for $\rho > 0.5$. This is due to the rearrangement of the partial contributions to dissimilarity from different scales $\mathcal{D}_{\rm xt}^k$ (see Supplemental Material~\cite{SM}) and the fact that we do not consider $k = 0$ step. It should be mentioned, that to get smooth graph with zero value at $\rho = 0.5$ we still have to consider sufficient number of time steps such that the dynamics at the critical point stops.   

The deeper analysis of the dissimilarity (see Fig.~\ref{fig:d_avg}) shows that the behaviour of $\mathcal{D}_{\rm xt}$ is similar to the fully spatial case $\mathcal{D}_{\rm x}$. Moreover, the slight differences occur only in active phase when the temporal part is high enough. Therefore, spatial patterns in digital representation provide most of meaningful information about the system. Nevertheless, having only the $\mathcal{D}_{\rm t}$ one can still define a phase transition analysing the derivative behaviour, as it was proposed in Ref.~\cite{PNASComplexity}.

\subsection*{Patterns of discrete time crystal}

The main focus in our paper is on the classification of dynamical phases realized in driven non-equilibrium systems of qubits. More specifically, we are interested in exploring  the properties of a periodically driven quantum system hosting the discrete time crystal (DTC) state. As it was shown in the previous works\cite{DTC1,DTC2,DTC3} DTC is formed as a result of  breaking discrete time translation symmetry in a Floquet system with time-periodic Hamiltonian and retaining memory of the initial state indefinitely. Experimentally, it can be realized in systems of superconducting qubits \cite{DTC_IBM, DTC_Google, Khemani}, but other platforms, for instance quantum-optical systems \cite{Fedorov} are also promising candidates for observing this non-trivial state. The main signature of DTC, the spatio-temporal order can be detected and probed with the combination of temporal autocorrelators and qubit-qubit spatial correlation functions. As it was demonstrated theoretically \cite{Khemani} and experimentally\cite{DTC_IBM, DTC_Google}, these correlation functions can be defined using the results (bitstrings) of the numerous projective measurements performed in a number of consecutive Floquet cycles during the evolution. In this work, by using the developed spatio-temporal dissimilarity measure we show that one can perform a complete low-level characterization of DTC with a limited number of bitstrings, directly without calculating correlation functions.

    \begin{figure}
		\centering
		\includegraphics[width=0.45\textwidth]{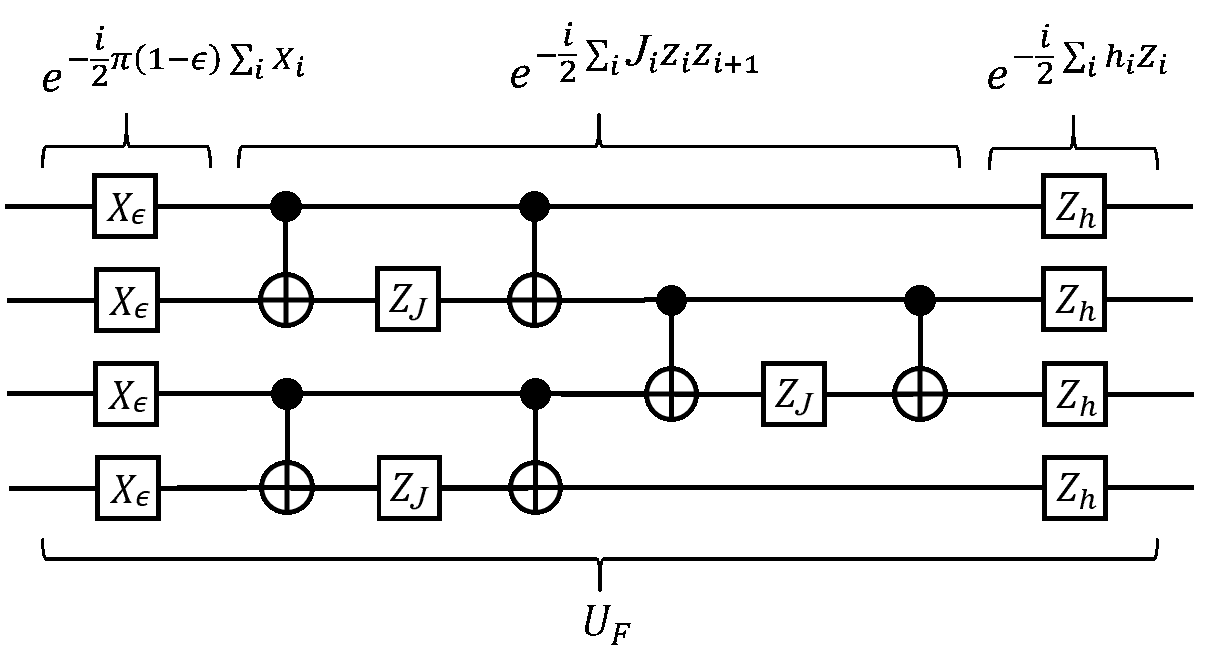}
		\caption{\label{fig:DTCcircuitInUniversalGates} Example of DTC quantum circuit for 4-qubit system where single-qubit rotational gates ($X_\epsilon=R_x((1-\epsilon)\pi)$, $Z_{J}=R_z(J_i)$, $Z_h=R_z(h_i)$) and CNOT-gate were used.}
	\end{figure}
    
Time evolution of the quantum system needed to stabilize non-equilibrium discrete time crystal state is described by the Floquet unitary $U_{\rm F}$ \cite{DTC_Google}:
\begin{multline} 
\label{U_F}
        U_{\rm F} = 
        e^{ \scalebox{1.0}{$-\frac{i}{2} \left(1-\epsilon\right) \pi \sum\limits_{i} X_i$} }   \\
        \cdot e^{\scalebox{1.0}{$-\frac{i}{2} \sum\limits_{i<j} J_{ij} Z_i Z_{j} -\frac{i}{2} \sum\limits_{i} h_i Z_i $ }}
\end{multline}
which is defined in term of two unitaries.
The first one corresponds to the imperfect global spin-flip (imperfect $\pi$-pulse). Here $X_i$ is the Pauli X-gate acting on the $i$-th qubit, $\epsilon\in[0;0.5]$ is the spin-flip imperfection parameter. The second unitary represents a combination of a random nearest-neighbor Ising interactions, $J_{ij}$ and a random longitudinal field, $h_i$. Here $Z_i$ is the Pauli Z-gate on the $i$-th qubit. Following to Ref.~\cite{DTC_Google} we use a trotterized time evolution to simulate the DTC with a quantum simulator. 

In Fig.~\ref{fig:DTCcircuitInUniversalGates} we show a fragment of the quantum circuit that corresponds to one trotter cycle for a 4-qubit system driven with Floquet unitary described above. It comprises rotation single-qubit gates and entangled CNOT gates. To construct a complete quantum circuit that imitates periodically driven quantum system for a given number of time steps we randomly choose  nearest neighbors interaction parameters $J_{ij}$ and local magnetic fields $h_i$ from the ranges $\left[-0.75\pi;-0.25\pi\right]$ and $\left[-\pi;\pi\right]$, respectively. These parameters are fixed for each quantum circuit instance independent on its depth. Wherein the imperfect spin-flip, $\epsilon$ plays a role of a global parameter whose value allows to choose a particular non-equilibrium phase. To be more concrete, $\epsilon$ is varied within [0, 0.5]. In our work we perform simulations of the DTC on the 16-qubit one-dimensional system with open boundaries.

\begin{figure}[!t]
	\centering
	\includegraphics[width=0.4\textwidth]{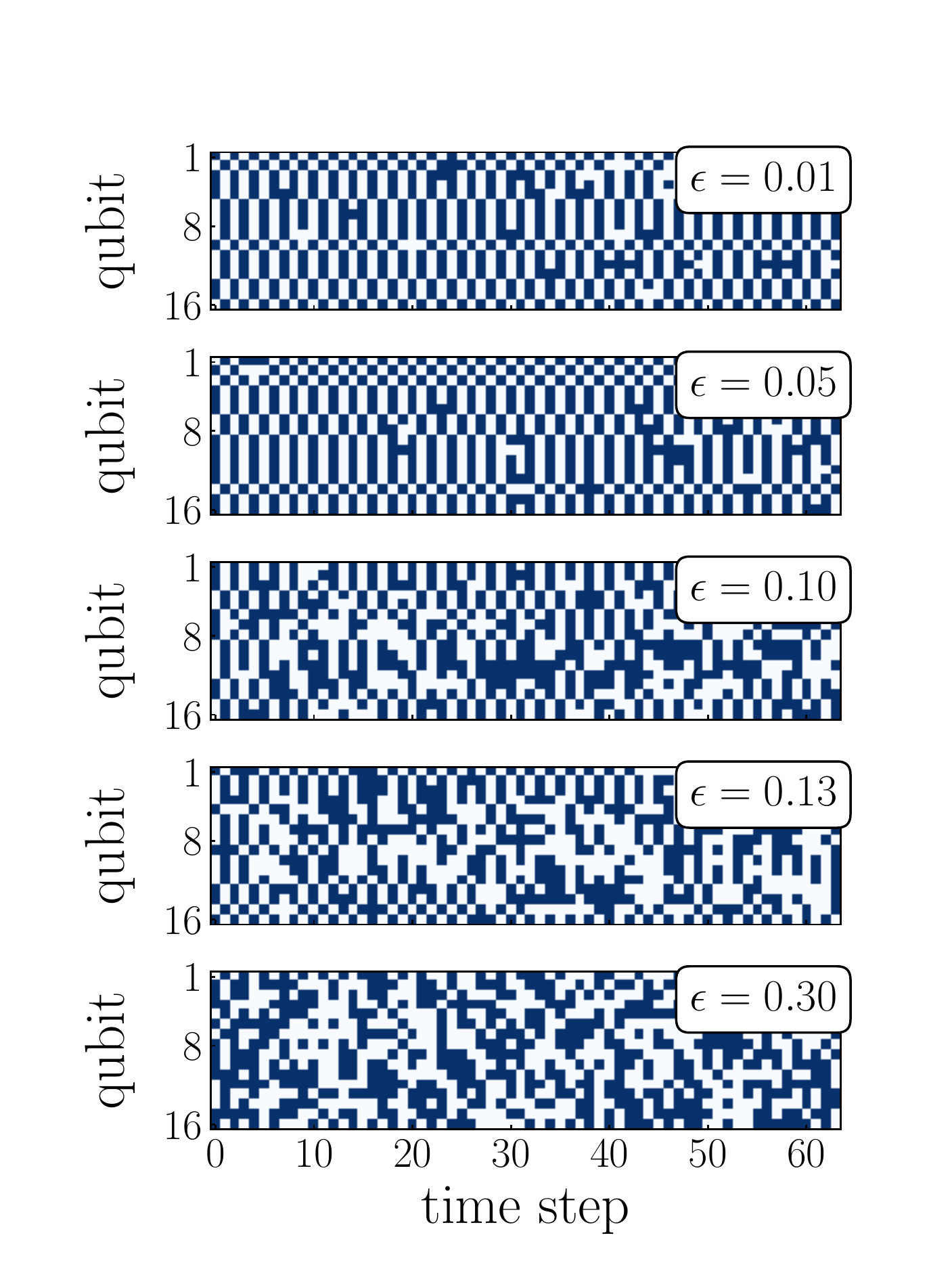}
	\caption{Examples of bitstring evolution within the time window [0, 63] for particular circuits constructed with five representative values for the $\epsilon$ parameter. \label{bitstrings}}
\end{figure}

While the previous works devoted to the DTC were mainly focused on the analysis of different forms of the spin polarization, here we show that the non-trivial state can be fully described at a more basic level of bitstrings. From 
Fig.~\ref{bitstrings} it follows that there are two different types of bitstring temporal evolution depending on the parameter of imperfect rotation. At small values of $\epsilon$ one can observe period-doubled oscillations between initial bitstring and its flipped counterpart at two consecutive time steps. At some times the projective measurement shows slightly modified bitstrings that differ from the initial one by a few bits. The latter means that quantum state becomes slightly delocalized in the Hilbert space with respect to the initial trivial state. At the same time, there are nearly perfect oscillations at the level spin polarization \cite{Khemani} within the DTC phase, when each spin is fully flipped. Thus, although from the point of view of structure, the quantum state changes during the evolution due to disorder, nevertheless it reproduces the period-doubled dynamics in terms of the correlation functions local in space, which is a signature of DTC.   

As $\epsilon$ increases, bit-string patterns are becoming more and more diverse over time, which means that different basis functions are involved and the delocalization of the quantum state increases. For $\epsilon = 0.3$ we observe a complete loss of information about the initial configuration.   

\begin{figure}[!b]
	\centering
	\includegraphics[width=0.35\textwidth]{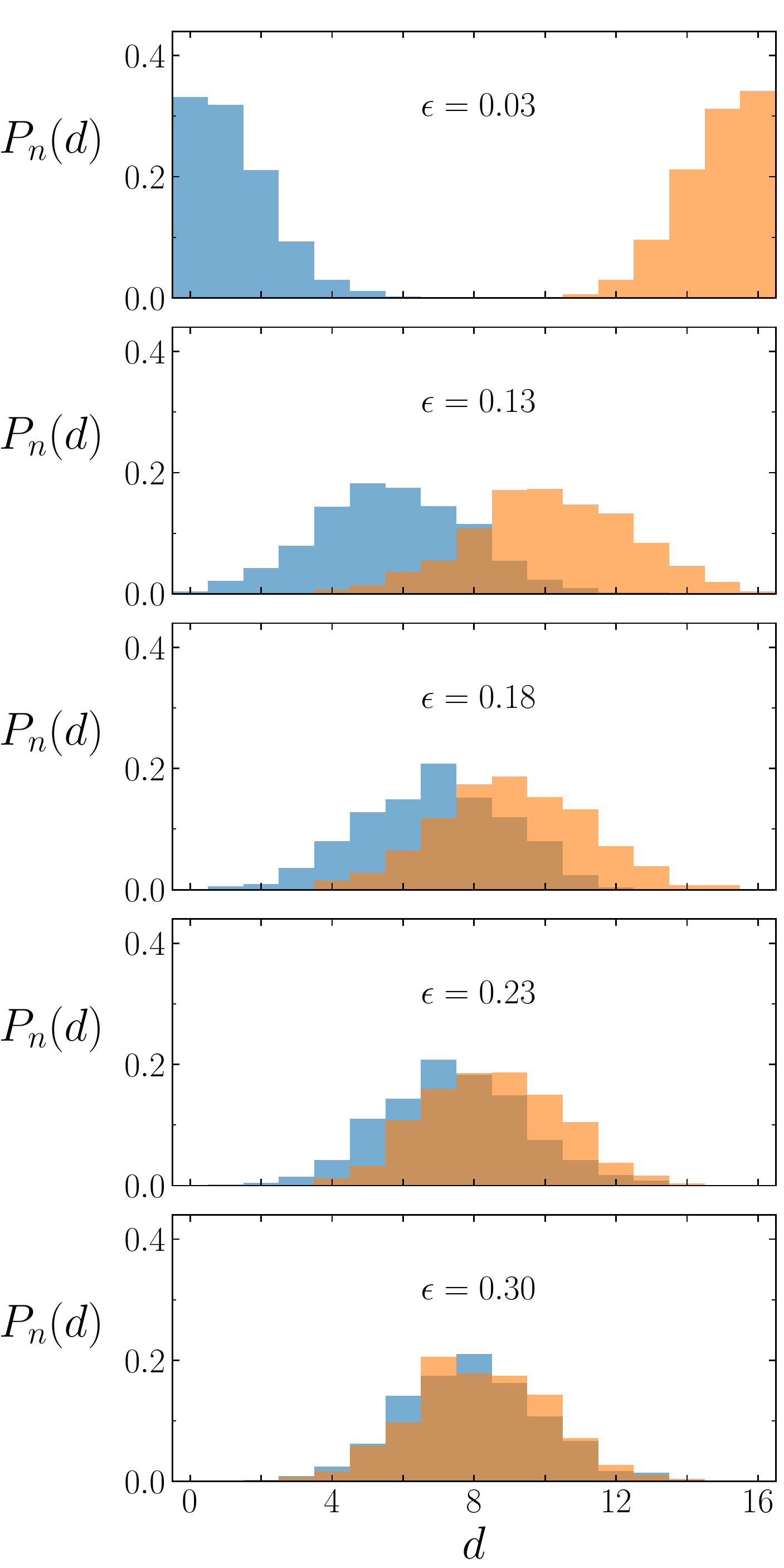}
	\caption{ Probability distribution of the Hamming distances $P_n(d)$ computed between the initial state and bitstrings measured at time steps 62 (blue) and 63 (orange) of the Floquet evolution. The histograms are obtained using 512 independent circuits.}
	\label{Hamming}
\end{figure}

To quantify the bitstring dynamics described above we employ the Hamming distance that measures the number of bits by which one bitstring differs from another and is defined as $d$. Firstly, the probability distribution of Hamming distances, $P_{n} (d)$ between the initial and time-evolved states is calculated. However, in contrast to  Ref.~\cite{Khemani} for our analysis we use a limited number of measurements, which does not allow to reconstruct the complete probability of the quantum system at each time step. Dependence of $P_{n}(d)$ on the parameter of imperfect rotation presented in Fig.~\ref{Hamming} unveils two boundary cases. The first one is observed at small $\epsilon$ and represents a signature of the DTC phase according to the results of Ref.~\cite{Khemani}. The probability distributions of the Hamming distances calculated between the initial state and two consecutive states obtained after a moderate number of time-evolution steps are localized at $d = 0$ and $d = 16$.

The distributions obtained with $\epsilon = 0.23$ and $\epsilon = 0.30$ (Fig.\ref{Hamming}) correspond to another distinct regime of the thermal phase for which $P(d)$ is centered at $d = 8$ for even and odd time steps. While DTC and thermal regimes of dynamics can be identified with the probability distribution of the Hamming distance, the transition between them on the level of the calculated $P(d)$ is smooth and does not show any feature that can be associated to the critical point. The presented distributions are not symmetrical with respect to $d = \frac{L}{2}$, since the number of bitstrings we used is much smaller than the total size of the Hilbert space for 16-qubit system. In Supplemental Material we complete the picture of the non-equilibrium dynamics in the considered quantum system by analyzing the behaviour of the Shannon and von Neumann entropies.

\begin{figure}[t!]
	\centering
	\includegraphics[width=0.45\textwidth]{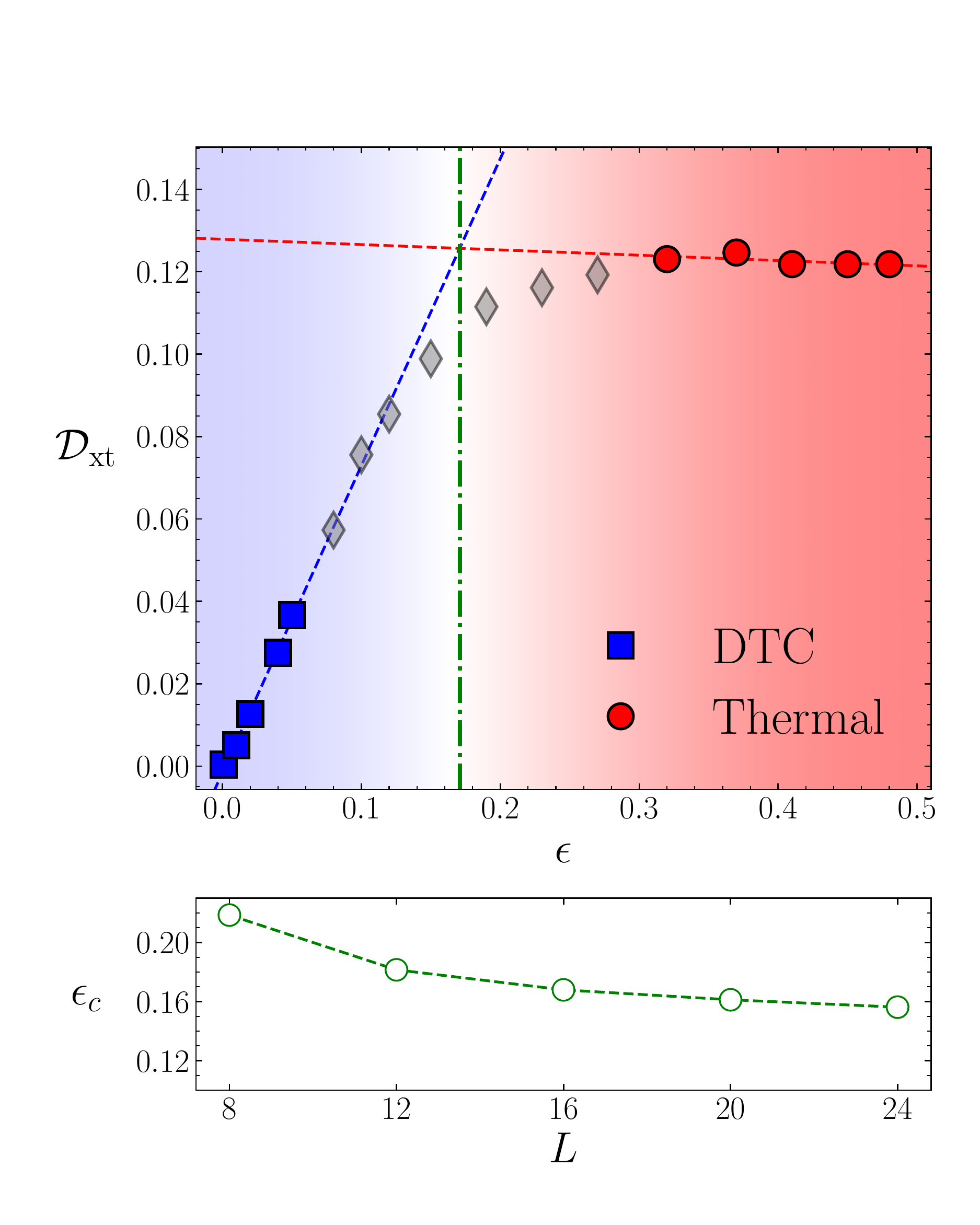}
	\caption{ (Top) Dependence of the spatio-temporal dissimilarity on the critical imperfect rotation parameter. Blue squares and red circles indicate the internal regions of the DTC and thermal phases, respectively. Dashed blue and red lines show the linear fittings for the corresponding points. Green dash-dotted line denotes the value of $\epsilon_c$ corresponding to the DTC-thermal transition. (Bottom) Dependence of the $\epsilon_c$ on the system size $L$. These results were obtained with 4096 ($L$=8), 6144 ($L$=12), 8192 ($L$=16), 4000 ($L$=20) and 4800 ($L$ =24) measurements. } \label{dissimilarity}
\end{figure}

As it was shown previously~\cite{Khemani}, the transition point between DTC and thermal phases can be defined with the Edwards-Anderson spin glass order parameter \cite{Anderson}, which represents a total spin-spin correlation function averaged over several time steps and different instances of disorder. In our work to explore a dynamical order and define critical value of $\epsilon$ we use the developed $\mathcal{D}_{\rm xt}$ measure that allows to define the critical point directly from the spatio-temporal representation of bitstrings. The key moment of our scheme is that at each time step of the quantum evolution we perform only one measurement of all the qubits and the resulting bitstrings obtained from several consecutive time steps are used to form qubit-time grid. For each value of $\epsilon$ the dissimilarity $\mathcal{D}_{\rm xt}$ was averaged over 512 realization of disorder (circuit instances) each of which assumes a random initial state. We have found that the minimal number of the time cycles required for calculating spatio-temporal dissimilarity in the case of DTC is equal to 16. Moreover, an increase in this number does not change the $\mathcal{D}_{\rm xt}$ estimates. In Supplemental Material we discuss different averaging schemes with a fixed disorder or a fixed initial state. 

The resulting dissimilarities are given in Fig.~\ref{dissimilarity}. One can clearly distinguish between discrete time crystal and thermal phases by analyzing the behaviour of the calculated $\mathcal{D}_{\rm xt} (\epsilon)$. In the DTC region the dissimilarity demonstrates a rapid growth with increasing $\epsilon$, which can be explained by the fact that the target wave function becomes more delocalized in the Hilbert space. In turn, in the thermal phase $\mathcal{D}_{\rm xt}$ is independent on this parameter and shows nearly zero slope. The critical value of the imperfect rotation parameter can be found as a crossing point of the corresponding linear fits \cite{glass1} of the results deep inside the DTC and thermal phases. Dependence of the $\epsilon_c$ on the system's size presented in Fig.\ref{dissimilarity} bottom shows the stabilization of the critical value for $L \geq 16$. The obtained $\epsilon_c\approx0.16$ is in good agreement with theoretical estimations based on the Edwards-Anderson order parameter\cite{Khemani} and lies within the range of experimental estimates~\cite{DTC_Google}.

The results presented in Fig.\ref{dissimilarity} were obtained for random initial states of zero entanglement, which is a standard setup widely used in the literature devoted to the DTC problem. In Supplemental Material we discuss the simulations of the DTC phase with different entangled initial states.

As it was shown in the classical case for one-dimensional CLG the exploration of the non-equilibrium critical behaviour with spatio-temporal dissimilarity can be successfully performed using much less data about the system in comparison with other measures. Thus, in the quantum case it is instructive to compare the total number of measurements required for characterization of DTC with $\mathcal{D}_{\rm xt}$ and standard order parameter of the spin glass state used in previous works \cite{Khemani}. To get one point on the graph presented in Fig.~\ref{dissimilarity} we use 8192 measurements (number of time steps $\times$ number of disorders). In turn, to estimate the glass order parameter for the given $\epsilon$ in the real experiment the authors of Ref.~\cite{DTC_Google} have performed 1600000 measurements which is mainly caused by noise and gates imperfections. In contrast, theoretical results on the correlation functions presented in the same work for the system of 16 qubits were obtained with 40000 measurements, which is still considerably larger than  those needed to calculate the dissimilarity.

\subsection*{Monitoring a quantum evolution}
The models of discrete time crystal and one-dimensional lattice gas we considered above feature single-regime dynamics for chosen parameters. This means that a dynamical process is fully associated to the particular non-equilibrium phase of the system and the calculated value of the spatio-temporal dissimilarity is weakly affected by the particular choice of the time window within the whole evolution. However, in general case a dynamical system out of equilibrium can pass different phases during its evolution, which requires developing methods for monitoring the system's state in time. To examine such multi-regime case with the spatio-temporal dissimilarity we consider a non-equilibrium system that was investigated in Ref.\onlinecite{transport} and features quantum spin transport.

\begin{figure}[b!]
	\centering
	\includegraphics[width=0.45\textwidth]{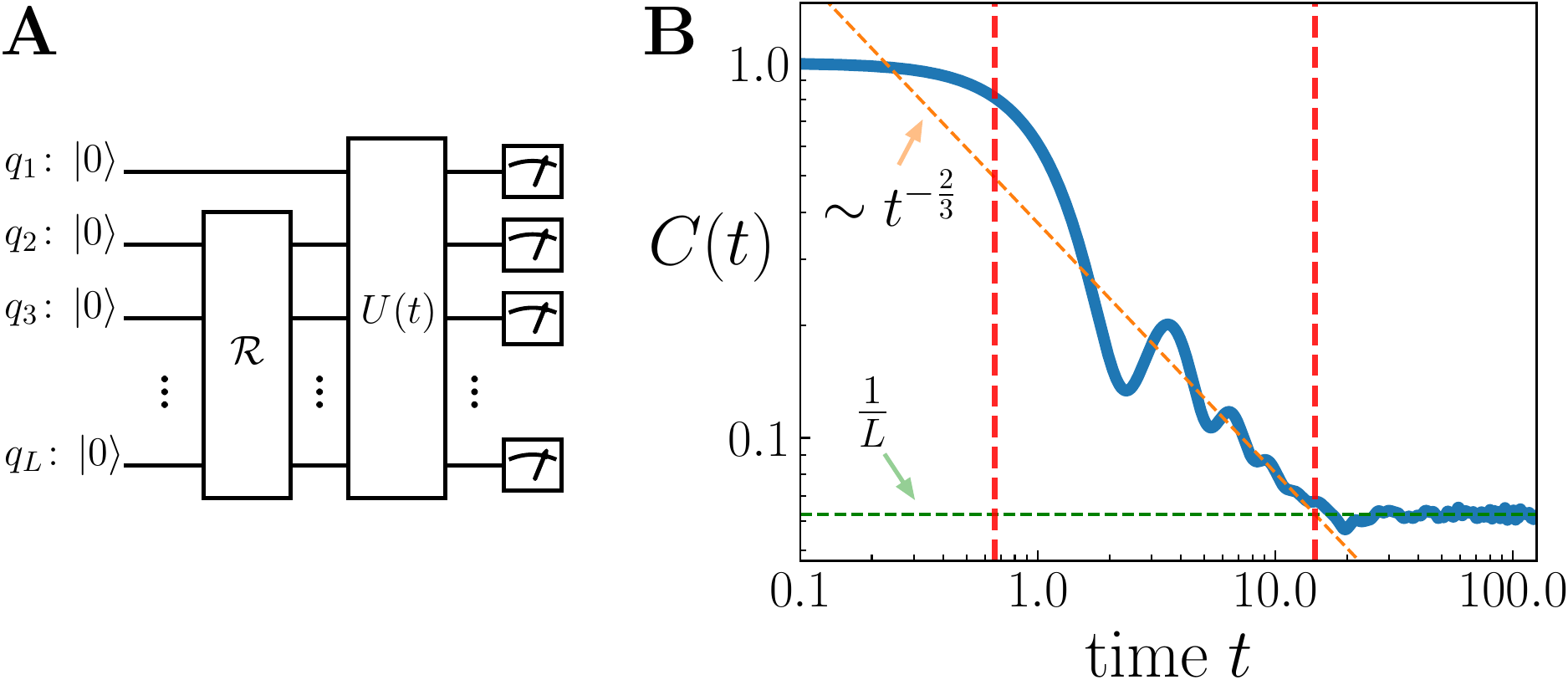}
	\caption{(\textbf{A}) Quantum circuit consisting of the block $\mathcal{R}$ that produces a random state for all the qubits except the reference one and block $U(t)$ that implements time evolution under the Heisenberg Hamiltonian for all the qubits. 
(\textbf B) Calculated equal-site correlation function for the reference qubit. These results were obtained for 16-qubits system with averaging over 128 experiments. Red lines denotes the transitions between different phases discussed in the text.}
    \label{Circuit_and_correlator}
\end{figure}

We simulate the many-body dynamics of 16-qubit system that is schematically visualized in Fig. \ref{Circuit_and_correlator} A. Initially, the system is prepared as a tensor product of the trivial $\Ket{0}$ state for the first qubit (reference qubit) and a Haar-random state, $\Ket{\mathcal{R}}$ for other qubits from 2 to 16 which can be considered as an environment for the reference qubit. The random quantum state $\Ket{\mathcal{R}}$ is approximated with shallow quantum circuits as described in \cite{transport}. Then, the states of the environment qubits and reference one are entangled within the unitary evolution under the Heisenberg Hamiltonian, $H = \sum_{ij} J_{ij}{\hat {\bf S}}_{i} {\hat {\bf S}}_{i}$. Here $J_{ij}$ is the antiferromagnetic exchange interaction between nearest neighbours in the spin chain that is characterized periodic boundary conditions. The corresponding time evolution operator is given by
\begin{eqnarray}
\label{evolu}
U(t) = (e^{-iH \delta t})^n \approx (e^{-iH_e \delta t} e^{-iH_o \delta t})^n + O(\delta t^2),
\end{eqnarray}
where $H_o$ ($H_e$) corresponds to odd (even) bonds of the spin chain, $\delta t = \frac{t}{n}$ is a Trotter time step and $n$ denotes the number of Trotter time steps. On the level of the quantum circuit realization of the time evolution \cite{transport}, each $e^{-iH_e \delta t}$ and $e^{-iH_o \delta t}$ is represented with CNOT and one-qubit gates. In turn, the Trotter time step is normally fixed at a small value to minimize the systematic error of the Trotter decomposition. Importantly, the authors of the work \cite{transport} have demonstrated that an accurate simulation of the quantum dynamics similar to that we consider can be performed by using the Trotter steps up to $\delta t = 1$. In our work we have simulated the quantum evolution by utilizing a dynamical Trotter time step that is defined as
$\delta t_{i} = t_0 \cdot \Delta d ^ i (\Delta d - 1)$, where $\Delta d = \left( \frac{T}{t_0} \right) ^ \frac{1}{N - 1}$ with initial time step $t_0$. We use the total number of times steps $N = 4096$, the maximum simulation time $T = 126$ and $t_0 = 0.01$.  Such parameters choice provides the sufficient number of the time steps for monitoring quantum dynamics with correlation functions and the spatio-temporal dissimilarity measure. It also guarantees the stability of the numerical scheme within the time window from 0 to $T$ in which dynamical phases of our interest are revealed. The Supplemental Material contains additional technical details about the scheme we use.

\begin{figure}[b!]
	\centering
	\includegraphics[width=0.45\textwidth]{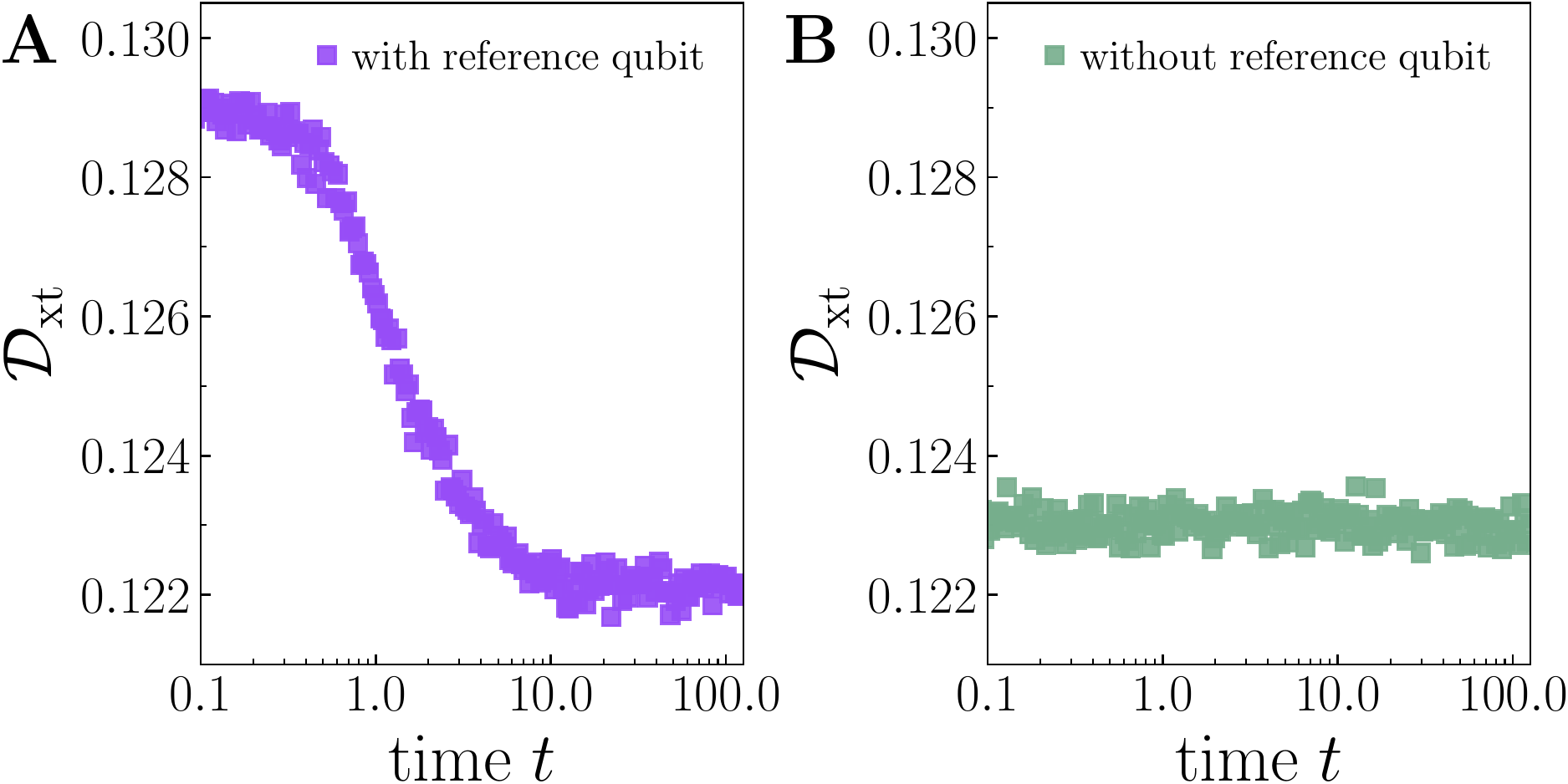}
	\caption{Calculated spatio-temporal dissimilarity as function of time that was obtained for 16-qubits system. 
    \textbf{A} The result was obtained with initial state that was prepared as a tensor product of the reference qubit and a Haar-random state.
    \textbf{B} An exclusively Haar-random state was used as initial state. 
    Both results were averaged over 12160 experiments.}
    \label{dissimilarity_XXZ}
\end{figure}

In previous works \cite{transport, transport2, transport3}, characterizing non-equilibrium dynamics was fulfilled by analyzing the expectation value of the reference qubit magnetization measured at different time moments $\Bra{\Psi (t)} S^z_1 \Ket{\Psi (t)}$. In the high temperature limit this magnetization was shown to be an approximation of the equal-site correlation function  
\begin{eqnarray}
C (t) = \frac{{\rm Tr} [{\hat S}^z_{1} (t) {\hat S}^z_1]}{2^L}.
\end{eqnarray}

Fig.\ref{Circuit_and_correlator} B gives the calculated magnetization of the reference qubit. The initial time evolution is characterized by the constant behaviour of the correlation function, which corresponds to the mean-free time \cite{transport2}.  In agreement with the results reported in work \cite{transport} we observe the hydrodynamic power-law tail $C(t) \sim t^{-2/3}$ at intermediate times in the range $t \in [5, 14.7]$. At the same time, for $t \in [0.7, 5]$ there are damped oscillations of the equal-site correlation function around the ideal $t^{-2/3}$ dependence. While the first critical time of 0.7 is defined from the minimum of the dissimilarity derivative discussed below, the second one corresponds to the transition to the constant behaviour of the correlation function, which is fully defined by the system size \cite{transport2}. Thus, the considered non-equilibrium model features different regimes at different time intervals that can be revealed with spin-spin correlation functions, which represents an important test for the spatio-temporal dissimilarity approach we propose in this work.  

While the characterization with the reference qubit magnetization described above can be considered as a local probing of the system's dynamics, the spatio-temporal dissimilarity measure assumes another level of the analysis. $\mathcal{D}_{\rm xt}$ allows to monitor the time evolution of the whole quantum system that includes the reference qubit and the quantum environment (2-16 qubits). For that the whole mesh of the 4096 time steps is divided into 256 non-overlapping time windows of 16 time steps of each. As in the case of the discrete time crystal we perform one measurement of each qubit in the system, which means that there is one bitstring per time step. Then we calculate $\mathcal{D}_{\rm xt}$ for each time window. The resulting time dependence of the dissimilarity is presented in Fig.\ref{dissimilarity_XXZ} A. 

Despite the environment subsystem initialized in an initial Haar-random state is much larger than that with single reference qubit prepared in trivial $\Ket{0}$ state, the resulting spatio-temporal dissimilarity reveals a distinct profile characterizing the quantum dynamics (Fig.\ref{dissimilarity_XXZ} A). Clearly, the environment introduces noise in the dissimilarity data which results in significant standard deviations of the calculated $\mathcal{D}_{\rm xt}$ when averaging over numerous experiments.  Nevertheless, the average dissimilarity profile is robust. Similar to the correlation function results (Fig.\ref{Circuit_and_correlator} B) there are two transitions in quantum system evolution around $t = 0.7$ and $t = 14.7$. The first one is defined as the derivative's extremum of the smoothed dissimilarity curve.  The calculated $\mathcal{D}_{\rm xt}$ is also characterized by the constant value for $t \ge 14.7$, which reproduces the behaviour of the magnetization of the reference qubit. To demonstrate such a dissimilarity profile is mainly related to the spin dynamics of the first qubit, we have performed the same calculations for 16-qubit system without the reference qubit. As follows from Fig.\ref{dissimilarity_XXZ} B the dissimilarity is characterized by a constant-like behaviour without any features. Thus, the spatio-temporal dissimilarity  allows one to detect and characterize a distinct dynamics producing in small parts of complex strongly entangled quantum systems. In the Supplemental Material we analyze a classical analog of the diffusion model we considered.

\section*{Conclusions}
We propose a flexible and simple approach for diagnosing non-equilibrium dynamics in classical and quantum systems. It allows identifying different dynamical regimes including hidden order out of equilibrium and accurate detecting critical parameters. Taking a time-resolved digital representation of a dynamical process as input, our method quantifies the variety or dissimilarity of spatio-temporal patterns by representing it as a single number. Thus, distinguishing different parts of the same process or different dynamical processes boils down to comparison of the corresponding numbers which considerably facilitates and accelerates characterization of non-equilibrium systems and constructing phase diagrams. 

By using the proposed spatio-temporal dissimilarity measure we have analyzed non-equilibrium dynamics of one-dimensional classical and quantum models. In classical case we show that  $\mathcal{D}_{\rm xt}$ reveals a hidden order in one-dimensional conserved lattice gas model for much smaller system's sizes than in the case of the compression measure. In turn, non-trivial dynamically driven discrete time crystal phase in a non-equilibrium quantum system can be fully described on the level of bitstrings without calculating standard correlation functions, which also substantially reduces the total number of the measurements. Another important application of our approach is demonstrated by the example of characterizing  spin transport in quantum systems. In this case one can explore the time evolution of a qubit subsystem was initially prepared in the localized state and embedded in the random fluctuating environment.   

In this work to extract information on a non-equilibrium one-dimensional system we explore two-dimensional spatio-temporal grid formed with digital representations of the system in question measured at different times. To consider higher dimensional physical systems, two different strategies can be adopted. The first one assumes using one-dimensional representation for spatial degrees of freedom of the high-dimensional system at each time step. As it was shown previously, in Ref.~\cite{Dissimilarity} such a dimensional reduction in space still allows an accurate description of equilibrium phase transitions in magnetic systems, however, it assumes testing different numeration schemes to order spatial degree of freedom and to mimic the topology of the initial system in one-dimension representation. Another possibility is to use high-dimensional spatio-temporal filters when calculating the dissimilarity, $\mathcal{D}_{\rm xt}$. For instance, in this work we have examined the approach with three-dimensional filters (2 spatial and 1 time dimensions) when describing classical spin dynamics presented in the Supplemental Material.

The dissimilarity method supports processing digital representations of completely different dynamical processes. For instance, in the main text we are concentrated on the characterization of the non-equilibrium quantum systems for which measurement's outputs are given in the binary format, bitstrings that consist of 0 and 1 (-1 and 1 in our calculations). However, we show that similar analysis can be done operating with a matrices whose elements are single floating-point numbers, when we consider the correlation functions as input data (see Supplemental Material). The ability to work with such vector fields paves the way to characterize the dynamics of off- and in-lattice models. From the perspective of real experiments one can calculate spatio-temporal dissimilarity directly for a stream of experimental images in RGB or other formats without mapping onto a mathematical model. This can be rationalized in non-scientific applications, for instance as a pre-processing scheme in automatic video content classification systems.

\section*{Acknowledgements}
The research funding from the Ministry of Science and Higher Education of the Russian Federation (Ural Federal University Program of Development within the Priority-2030 Program) is gratefully acknowledged. Quantum simulations were performed on the Uran supercomputer at the IMM UB RAS.

\end{document}